\documentclass[twocolumn,showpacs,preprintnumbers,prl,nofootinbib]{revtex4-1}
\pdfoutput=1
\usepackage{graphicx,epsfig,amsmath,amssymb,bm,slashed,wasysym,multirow}
\usepackage{color}
\usepackage{ifthen} 
\newboolean{pdflatex}
\setboolean{pdflatex}{True} 

\newboolean{articletitles}
\setboolean{articletitles}{true} 

\newboolean{uprightparticles}
\setboolean{uprightparticles}{false} 

\newboolean{inbibliography}
\setboolean{inbibliography}{false} 

\def \beq{\begin{equation}}
\def \eeq{\end{equation}}
\def \beqa{\begin{eqnarray}}
\def \eeqa{\end{eqnarray}}

\makeatletter
\DeclareRobustCommand*{\bfseries}{%
  \not@math@alphabet\bfseries\mathbf
  \fontseries\bfdefault\selectfont
  \boldmath
}
\makeatother

\usepackage{mciteplus}
\begin{document}

\preprint{IPPP/15/15}
\preprint{DCPT/15/30}
\title{The leptonic $W^{\pm}$ boson asymmetry in association with jets at LHCb, and PDF constraints at large-$x$.}

\author{Stephen Farry$^a$, and Rhorry Gauld$^b$}

\affiliation{
${}^a$\,Department of Physics, \\University of Liverpool, L69 7ZE Liverpool, UK,\\
${}^b$\,Institute for Particle Physics Phenomenology, University of Durham, DH1 3LE Durham, UK,\\
}

\date{\today}

\begin{abstract}
The unique forward LHCb acceptance opens the possibility for performing precision Standard Model (SM) processes in a kinematical regime beyond the reach of both ATLAS and CMS experiments. In this article we discuss the feasibility of performing $W^{\pm}j$ measurements with 14 TeV LHCb data, and find that the leptonic asymmetry at high pseudorapidity, with appropriate cuts on the associated jet, can provide a factor of three reduction on the $d$ quark PDF uncertainty in a region of Bjorken $x > 0.5$.
\end{abstract}

\maketitle

\section{Introduction}
With the large amount of data expected in Run II-A and Run II-B at the LHC, corresponding to $pp$ collisions at a centre-of-mass (COM) energy of 13 and 14~TeV respectively, precision Standard Model (SM) measurements and beyond-SM searches will become possible in kinematic regimes which have so far been statistically inaccessible. In both cases, it is necessary to reduce theoretical uncertainties to improve the sensitivity of searches and precision SM measurements.

One of the major uncertainties for many measurements at the LHC are those associated to the modelling of initial state partons. Fully describing the inelastic scattering of two hadrons is complicated due to the fact that the colliding hadrons are strongly-bound composite objects of interacting quarks and gluons. This complication has previously been overcome by factorising the full scattering process into both short-distance and long-distance effects~\cite{Drell:1970wh}. The short-distance effect describes a partonic interaction occurring at large momentum exchange scale $Q$ while the long-distance effects describe the evolution of partons within the incoming protons from a low energy scale $Q_0$ to the scale of the short-distance interaction $Q$. Although the partonic evolution can be computed perturbatively, as $Q_0$ is chosen to be well above the scale $\Lambda_{\rm{QCD}}$ at which the strong interactions are non-perturbative, the distribution of the partons within the hadron at $Q_0$ is a priori unknown. Rather, these parton distribution functions (PDFs) are extracted experimentally, based on a global analysis of data of many different measurements performed at many different energy scales. Several groups now provide modern PDF sets obtained from global fits to a large sample of data sets: ABM~\cite{Alekhin:2013nda}; CTEQ-TEA~\cite{Lai:2010vv}; CTEQ-JLAB~\cite{Owens:2012bv}; HERAPDF~\cite{CooperSarkar:2011aa}; MMHT~\cite{Harland-Lang:2014zoa} and NNPDF~\cite{Ball:2014uwa}. However, beyond-SM searches and precision measurements of SM processes are often limited by PDF uncertainties associated to the extraction of the PDFs from the global analysis. Improving the constraints on these PDFs will enhance our ability to test the SM and to look for new physics signals which may be kinematically accessible in Run II.

Due to the forward acceptance of LHCb, measurements of SM processes naturally provide information on the colliding partons which simultaneously carry a relatively large and small momentum fraction of the incoming protons momentum. As the probability of finding a parton declines with increasing $x$, measurements in which both partons are at large-$x$ are less likely than those with only one parton at large-$x$. Consequently, measurements at LHCb provide the best environment for constraining the large-$x$ region of PDFs. Indeed, several recent fits already include Run I LHCb data from various electroweak measurements~\cite{Aaij:2012mda,LHCb-CONF-2013-007,Aaij:2012vn}, which has improved the uncertainties of the $u$ and $d$ distributions at large-$x$. In this paper we suggest that measurements of $W^{\pm}$ production in association with jets within the LHCb acceptance can be used to constrain these PDFs at extremely large-$x$. 
In this letter we discuss the feasibility of $W^{\pm}j$ cross section and asymmetry measurements with Run II-B data and the expected improvement in the $d$-quark PDF at large-$x$. Further to this, we also demonstrate that requiring a significant $p_T$ cut on the reconstructed jet provides sensitivity to values $x \gtrsim 0.5$. Such data is of general importance to global fits as the available data in this region from deep-inelastic lepton-hadron scattering is statistically limited.

\section{Cross section predictions}
The goal of this article is to evaluate the potential sensitivity of $Wj$ measurements at LHCb to PDFs at large-$x$. The presence of a reconstructed jet in the final state acts to increase the partonic COM energy, predominantly providing sensitivity to the PDFs of proton travelling in the direction of the LHCb detector at large-$x$. A significant $p_T$ cut on the reconstructed jet is therefore vital for improving the sensitivity to PDFs. To avoid otherwise overwhelming QCD backgrounds, only the leptonic final state $pp \rightarrow l \nu_l$j is considered, and the provided differential cross sections correspond to only a single lepton flavour final state. Importantly, as the LHCb acceptance is limited to the forward-region, fully reconstructing the $W$ boson using transverse momentum conservation is not possible, due the ambiguity that missing transverse momentum may be due to undetected neutrinos or visible particles which are not within the detector acceptance.

The generation of the $Wj$ events is performed with \texttt{MadGraph5\_aMC@NLO}~\cite{Alwall:2014hca} subsequently showered with \texttt{Pythia8}~\cite{Sjostrand:2014zea}, achieving NLO+LL accuracy. The NLO NNPDF3.0 PDFs $\alpha_s(m_Z)$ = 0.118~\cite{Ball:2014uwa} are used for the generation of the hard process, and the default \texttt{Pythia8} \texttt{Monash 2013} tune~\cite{Skands:2014pea} is used. As constraints at large-$x$ are also of interest to the fitting of nuclear corrections to lepton- or proton-deuteron scattering processes, we also provide comparisons using the NLO CJ12 PDFs~\cite{Owens:2012bv}.

Charged leptons are required to have a $p_T$ greater than 20~GeV, and to be within the LHCb acceptance $2.0 < \eta_l < 4.5$.  To coincide with previous LHCb measurements which involve jet reconstruction~\cite{Aaij:2013nxa}, jet reconstruction is performed using the \texttt{FASTJET3} software~\cite{Cacciari:2011ma} with the anti-$k_t$ algorithm~\cite{Cacciari:2008gp} and a distance parameter of $R$ = 0.5. The jets are also required to be within the acceptance $2.0 < \eta_{\rm{jet}} < 4.5$. In the event selection, at least one reconstructed jet within the acceptance must have a $p_T$ greater than $p_T^{\rm{cut}}$ (which we vary), and an isolation criterion is placed on the lepton such that $\Delta R (l^{\pm}\,,\mathrm{jet}) \geq R$. Sub-leading jets are only reconstructed if they have a $p_T$ greater than 10~GeV.

\begin{figure}[ht!]
\begin{center}
\makebox{\includegraphics[width=1.0\columnwidth]{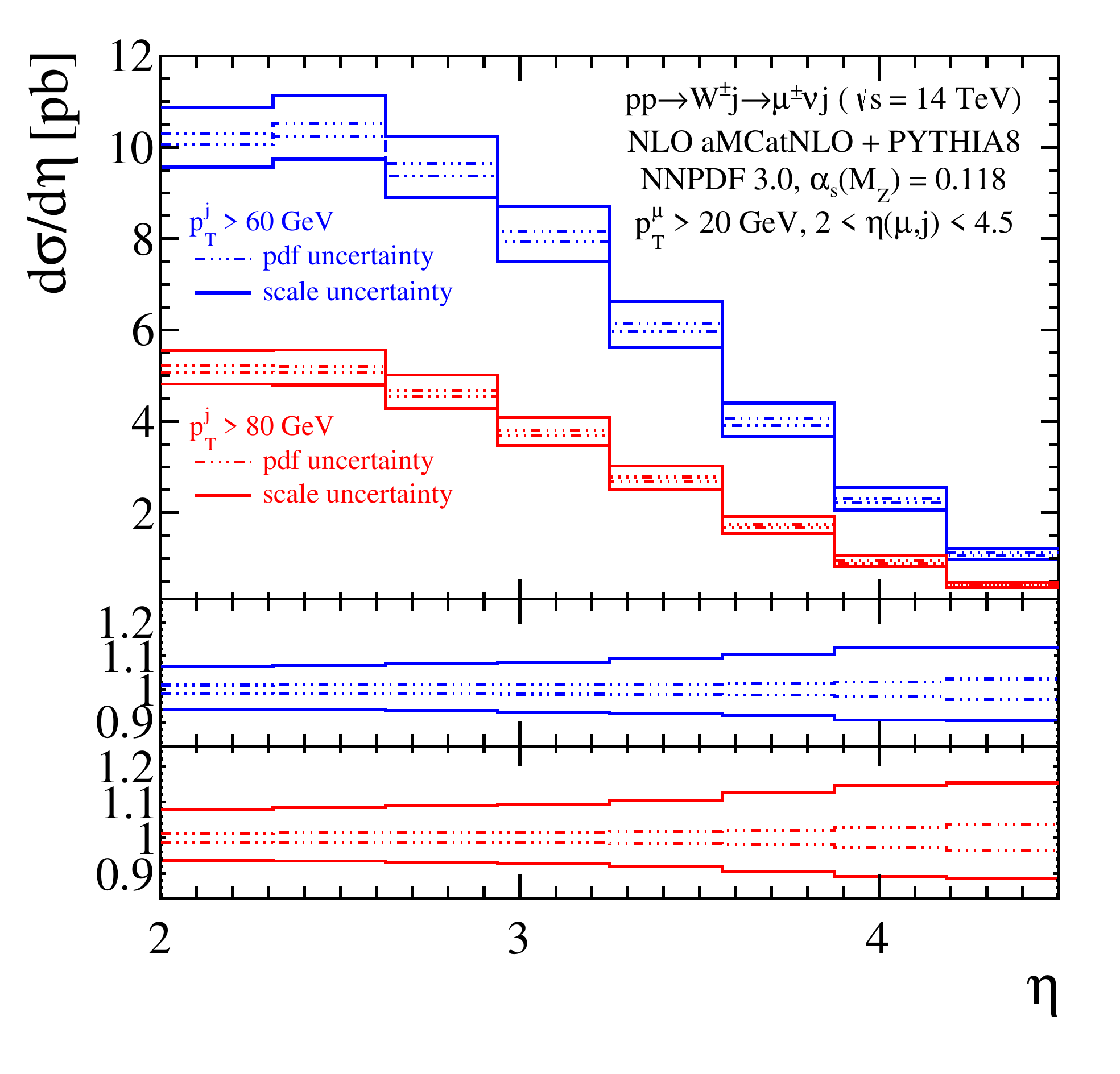}}
\end{center}
\vspace{-5mm}
\caption{Differential cross section with respect to charged lepton pseudorapidity. The sum of $W^{\pm}$j events are considered, and a single leptonic branching fraction has been applied. Distributions are shown for $p_{\rm T}^{cut}$ values of 60 (blue) and 80 (red) GeV where scale (solid) and PDF (dashed) uncertainties are displayed. For clarity, the relative uncertainties for the two sources are shown separately below the figure.}
\label{Differential}
\end{figure}

In Figure~\ref{Differential}, the differential cross section with respect to the lepton pseudorapidity is shown for $p_T^{\rm{cut}}$ = 60 and 80~GeV. 
A scale uncertainty is computed by varying factorisation ($\mu_F$) and renormalisation ($\mu_R$) sales independently by a factor of two around the nominal scale ($\mu_0$) with the constraint $1/2 < \mu_F/\mu_R < 2$. The nominal scale is set dynamically on an event-by-event basis to $\mu_0 = (\sum_i p_T^i)/2$, where $i$ corresponds to the $W$ boson and jets. The shown PDF uncertainty is the 68\% confidence interval. It is worth noting that the jet passing the $p_T^{\rm{cut}}$ within the LHCb acceptance is the leading jet for $\simeq$ 0.95 events, suggesting that the generation of $W^{\pm}+$1j at NLO is sufficient for the considered final state. The scale uncertainties (6-16)\% are dominant across the considered phase space, however the PDF uncertainties become significant for large values of lepton pseudorapidity. For the extreme lepton pseudorapidity bins ($3.9 < \eta_l < 4.5$), the average $< x_1 >$ values are 0.40, and 0.43 for $p_T^{\rm{cut}}$ = 60, 80~GeV respectively. For the considered PDF set, the shown uncertainty predominantly arises from lack of constraints on the $d$-quark PDF for large-$x$ and $Q^2$ values.

\begin{figure}[ht!]
\begin{center}
\makebox{\includegraphics[width=0.9\columnwidth]{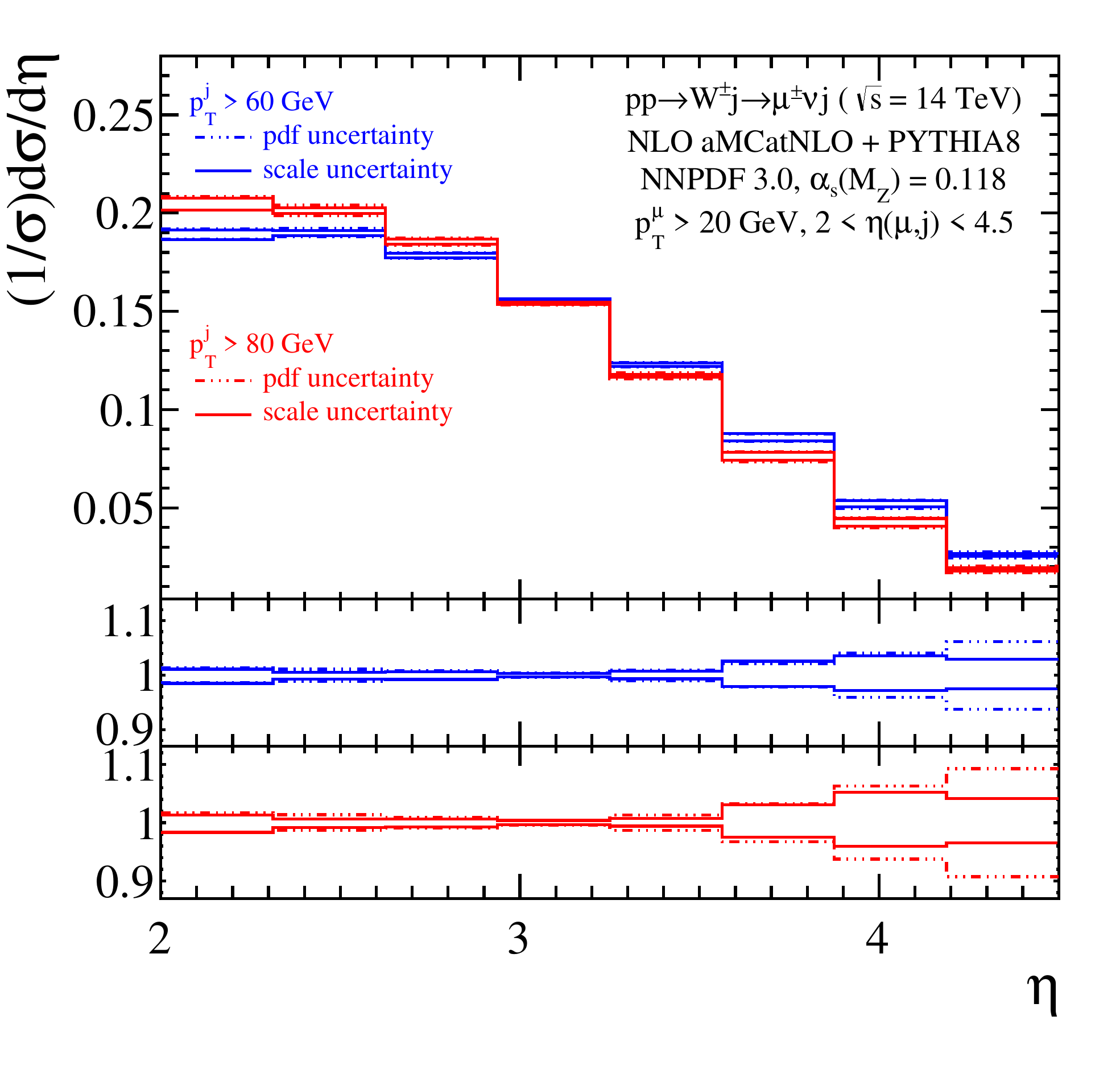}}
\end{center}
\vspace{-5mm}
\caption{Normalised differential cross section for the $W^-$j subprocess with respect to charged lepton pseudorapidity for $p_{\rm T}^{cut}$ values of 60 (blue) and 80 (red) GeV. The cross section is normalised with respect to the fiducial cross section. Both scale (solid) and PDF (dashed) uncertainties are displayed.}
\label{Normalised}
\end{figure}

In Figure~\ref{Normalised}, the normalised differential cross section for the $W^-j$ subprocess is shown. The distribution is normalised with respect to the differential distribution integrated over the experimentally accessible region corresponding to $\eta_l \in [2.0-4.5]$ --- referred to as the fiducial cross section. As less stringent constraints are present for the $d$-quark distribution compared to the $u$-quark, we have chosen to present only the $W^-j$ subprocess as the corresponding PDF uncertainties are largest.
By normalising the distribution, the renormalisation scale uncertainty is reduced as the leading contribution to this uncertainty, through the running of $\alpha_s(Q^2)$, partially cancels for this observable. For the pseudorapidity range $\eta \in [2.6-3.3]$, the average $Q^2$ of these bins is similar to the fiducial cross section which normalises the distribution, and causes the partial cancellation. The PDF uncertainties are less sensitive to this normalisation in the high and low-pseudorapidity bins as they depend on both $x$ and $Q^2$, and the $x$ dependence of the fiducial cross section and the high/low pseudorapidity bins is less correlated than the $Q^2$ dependence.

\section{Asymmetry predictions}
\label{sec:asymm}
Beyond normalised differential cross sections, a measurement of the charge asymmetry as a function of the lepton pseudorapidity is extremely well motivated. This observable is robust with respect to higher-order corrections as they do not differentiate between $W^+$j and $W^-$j subprocesses. The scale uncertainties only contribute through factorisation dependence, which can affect the $u/d$-quark ratio or the quark-antiquark PDF at a given value of $x$ and $Q^2$, both of which alter the asymmetry. 
Experimentally, this observable is also well motivated since many systematic uncertainties cancel. 
While not explicitly considered as a source of uncertainty, it has been shown~\cite{Corke:2010zj} that the choice of parton shower tune can also have an effect on the jet $p_{\rm T}$ distribution. Asymmetry predictions obtained alternatively using the older \texttt{Tune1} and \texttt{Tune4Cx} tunes of \texttt{Pythia8} show deviations of the asymmetry predictions on the order of the scale uncertainties.

\begin{figure}[ht!]
\begin{center}
\makebox{\includegraphics[width=1.0\columnwidth]{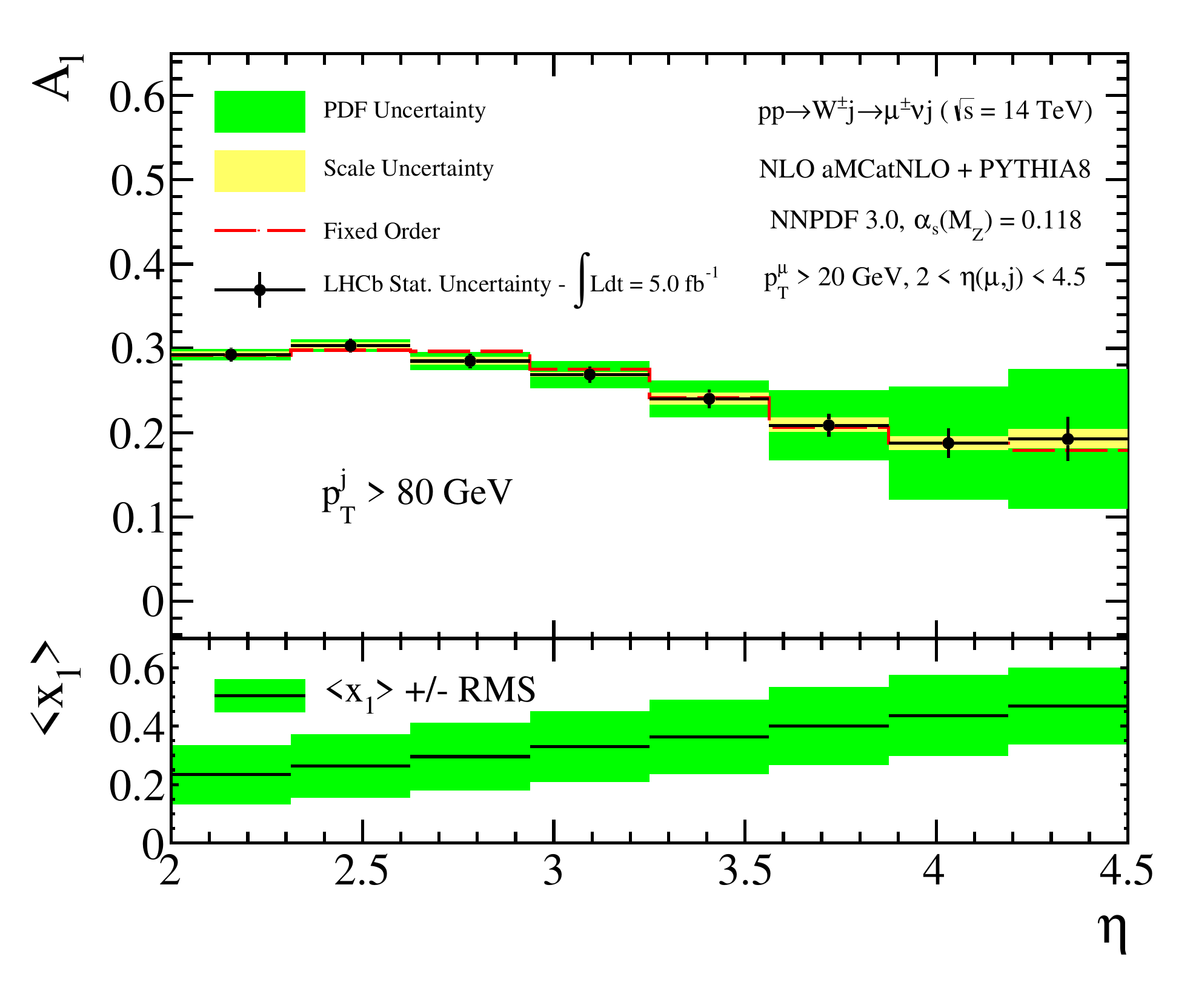}}
\end{center}
\vspace{-5mm}
\caption{The charged lepton asymmetry for selected events containing a jet with a $p_{\rm T}$ of greater than 80 GeV. Both scale (yellow) and PDF (green) uncertainties are shown, as well as the projected statistical uncertainty on measurements at LHCb with 5 fb$^{-1}$ of data. The mean value of $x_1$ is shown below with the band representing the root mean square of the distribution.}
\label{Asymmetry1}
\end{figure}

In Figure~\ref{Asymmetry1}, the differential asymmetry is shown for the choice $p_T^{\rm{cut}}$ = 80~GeV. The expected statistical precision after 5~fb$^{-1}$ of data taking is also shown and is obtained by scaling the expected yield of events by the reconstruction efficiencies reported in existing LHCb measurements. In particular, muon and jet reconstruction efficiencies of 0.75 and 0.95 respectively are determined in measurements of inclusive $W^{\pm}$ production~\cite{Aaij:2014wba} and $Z$ production in association with jets~\cite{Aaij:2013nxa} and are used in this study. In addition to the NLO+PS prediction, the corresponding central fixed-order prediction is also included showing satisfactory agreement. 

Noticeably, the lepton asymmetry tends to decrease beyond $\eta_l > 2.6$. Although the $W$ boson asymmetry increases at large rapidities, as the $u/d$-quark ratio and the quark-antiquark PDF asymmetry increase at large-$x$, the fraction of $W$ bosons produced in a left-handed polarisation also increases~\cite{Stirling:2012zt,Bern:2011ie}. As a consequence, the lepton asymmetry is reduced as the charged lepton from the polarised $W^{+(-)}$ boson decay prefers to be produced at lower(higher) rapidity with respect to the $W$ boson rapidity.

\begin{figure}[ht!]
\begin{center}
\makebox{\includegraphics[width=1.0\columnwidth]{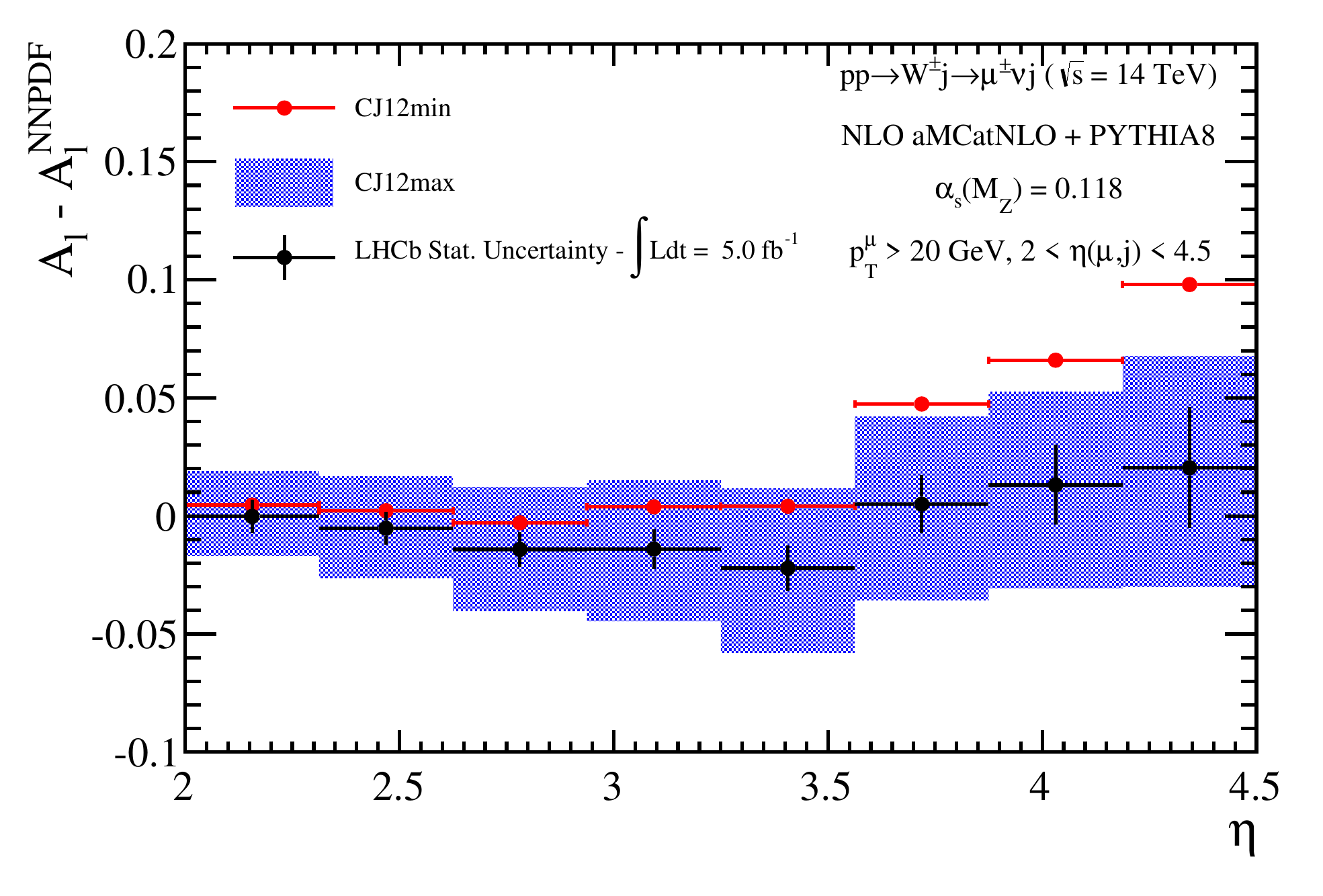}}
\end{center}
\vspace{-5mm}
\caption{The relative difference of the predicted charged lepton asymmetry with respect to the NNPDF prediction of Figure \ref{Asymmetry1} is shown for the CJ12min (red) and CJ12max(blue) PDF set. The PDF uncertainty is shown for the CJ12max PDF set. The statistical precision of the LHCb dataset with 5 fb$^{-1}$ is also shown, where the central value is fixed to that of the CJ12max prediction.}
\label{Asymmetry3}
\end{figure}

The relative difference of the predicted asymmetry with respect to NNPDF is shown for both CJ12min and CJ12max PDF sets in Figure~\ref{Asymmetry3}. 
The statistical uncertainties from Figure~\ref{Asymmetry1} are now centred on the CJ12max prediction. For CJ12max, the PDF uncertainty is explicitly shown and corresponds to the 1$\sigma$ Hessian uncertainty with a tolerance factor, $T=10$. While not explicitly displayed, a similar size uncertainty is also found for the CJ12min set. At high pseudorapidity the CJ12min and CJ12max predictions become differentiable. The difference between the shown CJ12 predictions is the assumption on the strength of nuclear corrections employed in the simultaneous fit of PDFs and nuclear corrections (required to model the deuteron bound state) in the CJ12 global analysis, which suggests that LHCb data can indirectly provide useful information on the strength of these corrections when included in such a fit.

\section{Estimated PDF uncertainty reduction} \label{Sec:Nuclear}
To quantify the sensitivity of the proposed asymmetry measurement, a reweighting of the NNPDF3.0 NLO replica PDFs is performed following the procedure outlined in~\cite{Ball:2010gb,Ball:2011gg}. Although no LHCb data is currently available for this observable, the reweighting procedure can be performed with respect to the central PDF prediction. This allows the reduction of the relative PDF uncertainty to be quantified when assuming different levels of experimental precision. 

This precision is defined as the relative uncertainty on the ratio of $W^+j$ to $W^-j$ production, $R_{Wj}$, which is related to the asymmetry, $A$, through the formula $A = (R_{Wj}-1)/(R_{Wj}+1)$. While the relative uncertainty on $A$ depends on the central value and is largest for asymmetries close to zero, the relative uncertainty on $R_{Wj}$ provides a more meaningful estimate of the experimental precision, and can be translated into an uncertainty on the asymmetry for any given central value.

%
%

In the measurement of inclusive $W$ production at LHCb~\cite{Aaij:2014wba}, the uncertainty on the ratio is dominated by the purity determination and is reported to vary between 1 and 3\% across the different pseudorapidity bins. The addition of a jet to the final state is not expected to inflate the uncertainty to any great extent, as the uncertainties related to jet reconstruction will also largely cancel in the ratio. While the finite jet energy resolution of the detector, reported to vary between 10-15\% in~\cite{Aaij:2013nxa}, will result in a migrations of events above and below the chosen jet $p_{\rm T}$ threshold, a smearing of the jet energy by this amount in the simulated samples indicates that the effect will be small. Consequently, the precision of the measurement is expected to be at the level of a few percent.

\begin{figure}[ht!]
\begin{center}
\makebox{\includegraphics[width=1.0\columnwidth]{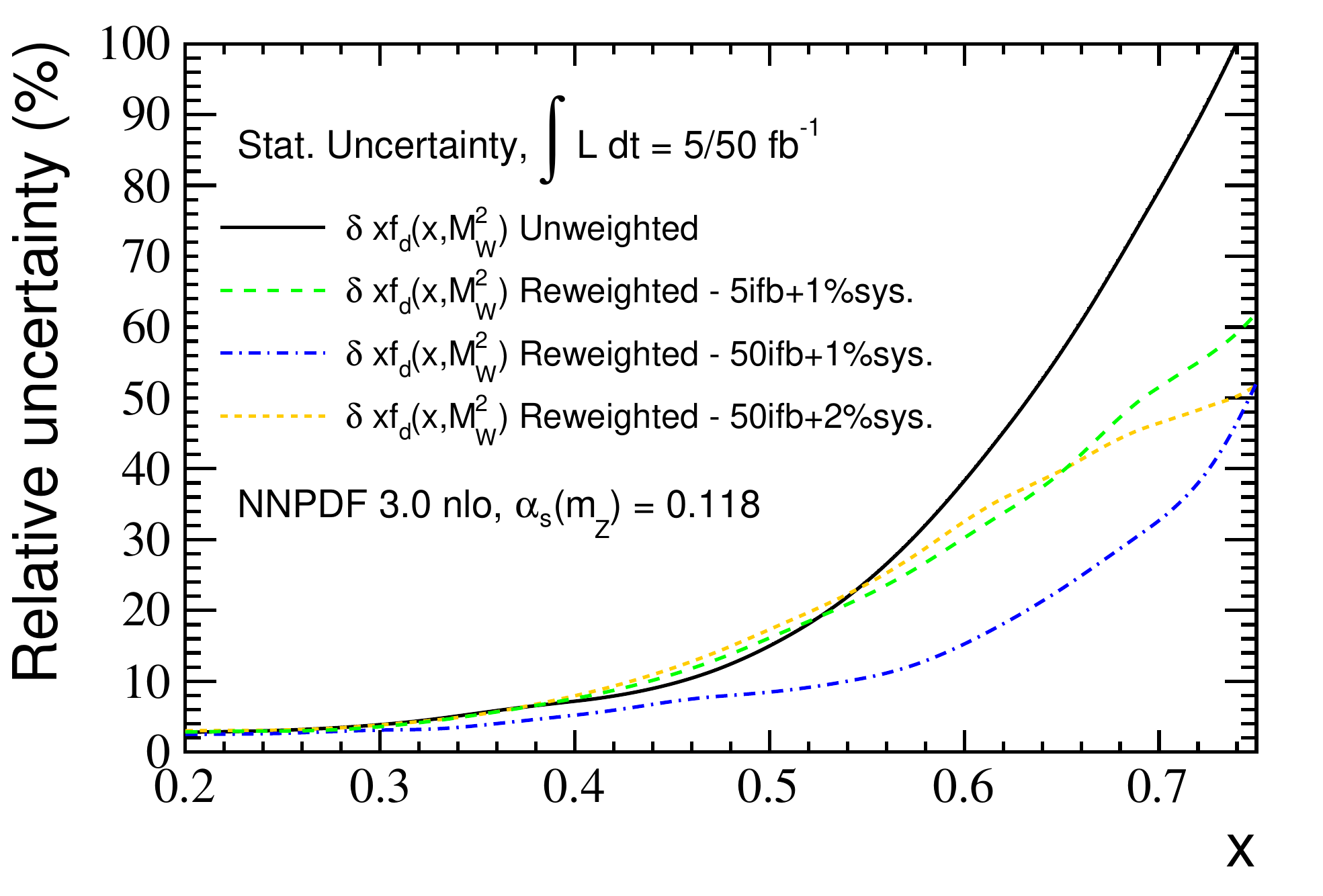}}
\end{center}
\vspace{-5mm}
\caption{Relative $1\sigma$ uncertainty of the evolved $d$-quark PDF. The reweighting is performed with respect to the central PDF prediction, and the sensitivity of different experimental precisions are also included.}
\label{Reweighted}
\end{figure}

The reweighting is performed assuming 5/50~fb$^{-1}$ of data is collected, and the impact on PDFs is studied if the additional systematic uncertainties can be controlled at the level of 1 or 2\%, which are added in quadrature with the statistical uncertainty in each bin. The sample with a $p_T$ threshold for the leading jet of 80~GeV (see Fig.~\ref{Asymmetry1}), which contains eight independent data points, is considered. 
To begin with, the probability distribution for the PDFs is provided by a finite ensemble of 100 replicas. The potential impact of this data is then estimated by computing a weight ($w_j$) for each of these replicas as detailed in~\cite{Ball:2010gb}, which is conditional on both the old data as well as the new data points which we are including. After reweighting the number of effective replicas can be quantified by using the Shannon entropy as:
\beq
N_{{\rm{eff}.}} = {\rm{exp}} \left[ \frac{1}{N} \sum_{j=1}^N w_j {\rm{Log}}(N/w_j)\right] \, .
\eeq
After reweighting assuming 50~fb$^{-1}$ of data, the original 100 replicas are reduced to 18 and 30 effective replicas when a systematic uncertainty of 1 and 2\% are assumed respectively. When performing reweighting with respect to the central value (as is done here), a reduction in the number of effective replicas indicates that the new data is highly constraining. The relative 68\% confidence interval of the original and reweighted evolved $d$-quark PDF are shown in Fig.~\ref{Reweighted}.

The $d$-quark PDF is already well constrained below $x$~=~0.5, however, significant improvements can be expected beyond this with the addition of LHCb data. As indicated in the lower plot of Fig.~\ref{Asymmetry1}, although the RMS of sampled $x$ values for high pseudorapidity events lies within the region $x \in [0.35,0.60]$, a significant fraction of events contribute both above and below this region. The observed PDF constraints for NNPDF3.0 are most significant at large-$x$ values. For example, at $x$ = 0.7, assuming a 1\% systematic and 50~fb$^{-1}$ of data, the relative reduction of the uncertainty of the reweighted with respect to unweighted PDF is $\simeq$ 60\%, corresponding to a factor of three reduction. Even when 5~fb$^{-1}$ of data is collected (expected after 1 year of data taking), the corresponding reduction is $\simeq$ 35\% indicating that early Run-II measurements will also be highly constraining.

It is worth noting that although the improvement in the PDFs has been quantified by only considering the total uncertainty on the $d$-quark PDF at $Q^2 = M_W^2$ ; correlated improvement in the other PDFs, which has not been quantified here, is also expected. For example, a reduction in the uncertainty for the $u$-quark PDF and $d/u$-quark PDF ratio at large-$x$ is expected, as well as for the PDFs at low-$x$ which also contribute the hard process.
The substantial reduction in the number of effective replicas in this analysis also suggests that a new global fit will be necessary to fully quantify the impact of this data when it becomes available. This can be seen explicitly in Fig.~\ref{Reweighted} in the region $x \in [0.4,0.5]$ where a marginal increase in the $d$-quark PDF is observed for $Q^2 = m_W^2$. This demonstrates that the weighted ensemble provides a poorer description of the underlying probability distribution.

\section{Discussion and conclusions}
We have proposed that measurements of the $Wj$ process with a $p_T$ threshold for the leading jet of $\simeq$ 60, 80~GeV should be performed at LHCb. Of particular interest is the lepton asymmetry which is both theoretically and experimentally clean, and sensitive to PDFs beyond $x$ = 0.5. In this work, we focussed on the potential of the $Wj$ final state without taking particular care to optimise kinematic cuts. We note that placing further restrictions on the final state, such as requiring the leading jet to also be forward ($\eta_j > 3.0$), or requiring the presence of a secondary jet within the acceptance with a moderate $p_T$ threshold ($p_T^{\rm{cut}} > 40$~GeV) further improves the sensitivity to yet larger values of $x$. Given that the proposed measurement is not anticipated to be statistically limited, these additional requirements should also be considered in future measurements. For example, a measurement of the lepton asymmetry differentially in $\eta_l$ as well as jet $p_T$ and/or $\eta_j$.

The proposed measurements have the potential to provide among the most stringent large-$x$ constraints from Run-II of the LHC. This is a consequence of the theoretical and experimental precision of the chosen variable, as well as the unique forward acceptance of the LHCb detector, which naturally provides sensitivity to large-$x$ values as compared to ATLAS or CMS detectors. Recent NNLO calculations~\cite{Boughezal:2015dva} will also increase the sensitivity of LHCb measurements to the large-$x$ gluon, which can then be used as a constraint as outlined in Ref.~\cite{Malik:2013kba}.

Precise differential $Wj$ cross section and asymmetry measurements are also important as a background to other interesting SM measurements which can be performed in the LHCb acceptance. For example, in other work~\cite{Kagan:2011yx,Gauld:2013aja,Gauld:2014pxa}, it has been suggested that top-quark pair cross section and angular production asymmetry measurements are possible at LHCb. In both cases, the statistically most promising final state is $t\bar{t}\rightarrow l bX$, for which the $Wj$ subprocess is the dominant background, and consequently measurements of $Wj$ production can significantly improve the precision of $t\bar{t}$ measurements at LHCb.
 
\section{Acknowledgements}
SF and RG acknowledge several useful discussions with Alberto Accardi which greatly improved the manuscript, and also thank Graeme Watt for his useful advice. RG also acknowledges support from the ERC Starting Grant PDF4BSM of Juan Rojo to attend the Parton Distributions for the LHC in Benasque where this work was preliminarily presented. The work of SF is supported by a fellowship from the Royal Commission for the Exhibition of 1851.

\bibliographystyle{BIB}
\bibliography{LearnHowToCiteRight}

\end{document}